\documentclass[12pt]{article}

\usepackage[utf8]{inputenc} % OVERLEAF DEFAULT
\usepackage{graphicx}

\usepackage[amssymb]{SIunits}

\title{A nonlinear, geometric Hall effect without magnetic field}

\author
{Nicholas B. Schade,$^{1\ast}$, David I. Schuster,$^{2}$ and Sidney R. Nagel$^{1}$\\
\\
\normalsize{$^{1}$Department of Physics and the James Franck and Enrico Fermi Institutes,}\\
\normalsize{ University of Chicago, Chicago, IL 60637, USA}\\
\normalsize{$^{2}$Department of Physics and the James Franck Institute,}\\
\normalsize{University of Chicago, Chicago, IL 60637, USA}\\
\\
\normalsize{$^\ast$To whom correspondence should be addressed; E-mail:  nschade@uchicago.edu.}
}

\date{}

\begin{document}

% Double-space the manuscript.

\baselineskip24pt

% Make the title.

\maketitle 

\begin{abstract}  The classical Hall effect, the traditional means of determining charge-carrier sign and density in a conductor, requires a magnetic field to produce transverse voltages across a current-carrying wire.  We show that along curved paths -- \textit{without} any magnetic field -- geometry alone  can produce nonlinear transverse potentials that reflect the charge-carrier sign and density.  We demonstrate this effect in curved graphene wires where the transverse potentials are consistent with the doping and change polarity as we switch the carrier sign.  In straight wires, we measure transverse potential fluctuations with random polarity demonstrating that the current follows a complex, tortuous path.  This geometrically-induced potential offers a sensitive characterization of inhomogeneous current flow in thin films.
\end{abstract}

In 1879, Edwin Hall discovered that a transverse potential appears across a current-carrying wire placed in a magnetic field~\cite{Hall1879}. Physics instructors traditionally teach the Hall effect when the magnetic field is first presented in introductory electricity and magnetism as the way to distinguish the sign of the charge carriers in a conductor~\cite{Purcell1985}. Indeed, the Hall effect is an efficient way to disentangle the role of electron and hole conduction in semiconductors~\cite{Busch1989} and is commonly used to measure the magnitude of magnetic fields~\cite{Ramsden2006}.  Here we present a different mechanism using geometry alone, without a magnetic field, to produce a transverse voltage that also reflects the sign and density of the charge carriers.  A current traveling through a curved wire must undergo centripetal acceleration to follow the curve.  This acceleration occurs due to electric fields from charges distributed along the wire edges; the direction of the field must change with the sign of the carriers.  No magnetic field is necessary.

The transverse voltage that we predict and measure is quadratic in the current. It is therefore distinct from many linear Hall effects such as the spin-~\cite{Dyakonov1971}, valley-~\cite{Mak2014}, and anomalous-~\cite{Nagaosa2010} Hall effects.  However, the geometric effect we describe should be contrasted with predictions~\cite{Sodemann2015} and recent measurements of a nonlinear Hall effect in nonmagnetic bilayer materials such as WTe$_{2}$~\cite{ma2019}.

By 1850, Kirchhoff realized that surface charge distributions are necessary simply to confine a current inside a wire~\cite{Kirchhoff1850}.  Subsequently, the role of surface charges has been investigated theoretically~\cite{Sommerfeld1952,Jackson1996,Muller2012} and experimentally using single-electron transistors~\cite{Yoo1997,Martin2008}.  The effect we describe is also related to how currents are confined within wires. Our experiments show that the effect can be used for sensitively probing non-homogeneous current flow in thin films as well as systems where magnetic fields do not occur such as the bulk of a superconductor.  While the effect is small, it is measurable even in bulk conductors. Using graphene wires to optimize the signal, we measure transverse potentials $\approx$ 0.5 mV.

In the conventional Hall effect, when an electric current $I$ of charge carriers $q$ passes through an applied magnetic field $\mathbf{B}$, the carriers experience a magnetic Lorentz force $\mathbf{F_B} = q \mathbf{v} \times \mathbf{B}$.  Charge accumulates at the wire edges, as illustrated in Fig.~1A, until the electric field caused by these charges cancels $\mathbf{F_B}$ to create a transverse potential difference:

\begin{equation}
    V_{\mathrm{Hall}} = \frac{I B}{t n q}.
    % NOTE:  There is a reference to this equation in the SI.
    \label{eq:V_Hall}
\end{equation}

\noindent Here $n$ is the charge-carrier density and $t$ the wire thickness along the direction of $\mathbf{B}$.  A measurement of $V_{\mathrm{Hall}}$ determines $n$ and the sign of the charge carriers.

\begin{figure}
    \centering
    \includegraphics{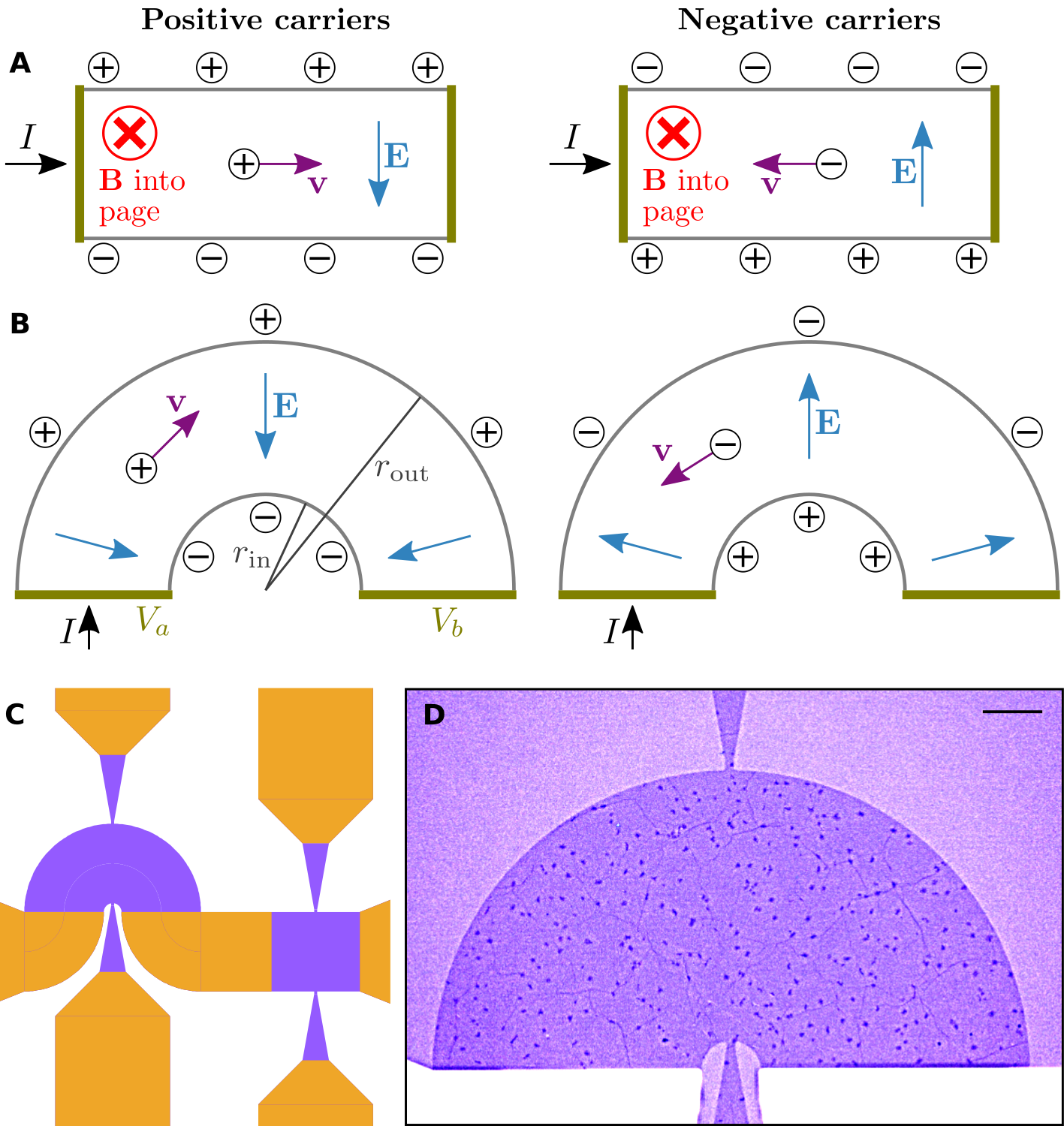}
    \caption{Surface charge distributions that produce transverse potentials.
    (\textbf{A}) In the classical Hall effect, the directions of current, $I$, and of the magnetic field, $\mathbf{B}$, determine the direction of the magnetic force.  The surface charges produce a transverse electric field $\mathbf{E}$.
    (\textbf{B}) In a curved wire without an applied magnetic field, the centripetal acceleration of the carriers is due to an electric force. Surface charges produce an electric field whose direction reveals the sign of the carriers.
    (\textbf{C}) Circuit for measurement of transverse potentials due to wire geometry.  Orange regions are metal and purple regions are exposed graphene.
    (\textbf{D}) Optical micrograph of the curved graphene wire in a completed device.  Graphene electrodes are visible in center at top and bottom.  Scale bar is 20 $\micro$m.
    }
    \label{fig:fig1}
\end{figure}

To show that magnetism is not necessary for producing transverse voltages sensitive to $n$ and the sign of $q$, we take advantage of surface charge distributions that are a result of geometry -- curvature in the path of the current -- rather than magnetic forces.  The surface charges create a transverse component of the electric field and exert a force on the current such that it follows the wire. For a wire in the shape of a circular arc, the transverse electric force must point radially towards the circle center to accelerate the carriers centripetally; this mandates that the direction of the radial electric field (and thus the potential difference between the wire edges) depends on the sign of the current carriers, as illustrated in Fig.~1B.  The polarity of this potential, analogous to the polarity of a Hall potential, reveals the sign of the carriers. 

Calculating this potential is straightforward. We assume the wire is a semicircular annulus with inner radius $r_{\mathrm{in}}$ and outer radius $r_{\mathrm{out}}$; the wire has a rectangular cross-section of width $w \equiv (r_{\mathrm{out}} - r_{\mathrm{in}})$ and thickness $t$. We force the radial edges of the half annulus to be equipotentials, as illustrated in Fig. 1B, with potential difference $\mathcal{E} = V_{a} - V_{b}$. Due to azimuthal symmetry, the charges on average follow semicircular trajectories. We assume that the conductivity $\sigma$ and carrier density $n$ are uniform and that all charge carriers have the same mass $m$ and charge $q$.

Unless the wire is a superconductor, the current density, $j(r)$, is equal to the longitudinal (azimuthal) component of the electric field, $E_\theta = \mathcal{E} /(\pi r)$, times the conductivity:  $j(r) = \sigma \mathcal{E}/(\pi r)$.  The transverse  (\textit {i.e.}, radial) component, $E_r$, provides the force necessary for the carriers to follow circular paths.

The local velocity $v$ of the carriers is proportional to the current density: $v = j/nq$ and the force for keeping these particles in a circular orbit of radius $r$ is $F = m v^2/r = E_r q$.  The radial electric field is thus
\begin{equation}
    E_r(r) = - \frac{dV(r)}{dr} = \frac{m \left[ j(r) \right]^2}{n^2 q^3 r} 
    \label{eq:E_field}
\end{equation}

\noindent where $V(r)$ is the electrostatic potential.  This can be integrated across the width of the wire, $w$,  to find $\Delta V_{\mathrm{geom}} \equiv V_{\mathrm{out}}-V_{\mathrm{in}}$: 

\begin{eqnarray}
    \Delta V_{\mathrm{geom}} & = & \left[
    \frac{r_{\mathrm{in}}^{-2} - r_{\mathrm{out}}^{-2}} {2 t^2 \left( \ln{ \left( r_{\mathrm{out}} / r_{\mathrm{in}} \right)} \right)^2} 
    \right] \left( \frac{m}{n^2 q^3} \right) I^2 
      \label{eq:V_trans} \\
    & \approx & \left[ \frac{1}{r_{\mathrm{in}} w t^2} \right] \left( \frac{m}{n^2 q^3} \right) I^2 = \left[ \frac{w}{r_{\mathrm{in}}} \right] \left( \frac{m}{n^2 q^3} \right) \left<j\right>^2  \ \ \ \ \ \ \ \ \ \mathrm{in \  limit} \ w \ll  r_{\mathrm{in}}.
      \label{eq:V_trans_approx}
\end{eqnarray}
% NOTE:  There is a reference to this equation in the SI.

\noindent The terms in square brackets depend only on the wire geometry; the approximations in Eq.~\ref{eq:V_trans_approx} are valid in the narrow-wire limit, $w \ll r_{\mathrm{in}}$, with $\left<j\right>=I/wt$ being the average current density.  As in the Hall effect, the potential $\Delta V_{\mathrm{geom}}$ is an odd power of the charge $q$, which means that the sign of $q$ can be determined.

There are two significant differences between this transverse potential and the Hall effect.  The first is that the mass of the carriers, $m$, enters into the expression for the potential difference.  Whether $m$ should be the bare or the effective mass of the carriers is an important open theoretical question.  The second difference is that $\Delta V_{\mathrm{geom}}$ is quadratic, rather than linear, in the current, $I$. 

To maximize the signal $\Delta V_{\mathrm{geom}}$, we need high current $I$ (or current density $j$) and low carrier density $n$.  It is advantageous to use a conductor whose charge-carrier sign and density can be modulated in order to check whether the signal and carriers change sign concurrently.  Monolayer graphene satisfies these conditions~\cite{Novoselov2004}. 

Our circuit, illustrated in Fig.~1C, consists of a curved graphene wire with measurement leads on either side, and a straight graphene wire as a control.  We use graphene grown by chemical vapor deposition and transferred by the manufacturer (Graphenea) to a doped silicon wafer with a 300 nm oxide gap.  We use photolithography, electron-beam evaporation, and plasma etching to pattern the graphene and to place Ti/Au contacts on it, as shown in Fig.~1D.  (See SI for detailed nanofabrication procedure.)  We control the sign and density of the carriers in the graphene by applying a back-gate voltage $V_{\mathrm{bg}}$ to the silicon.  As initially fabricated, the samples are highly doped; we current-anneal \cite{Moser2007} them to cross the Dirac point at $V_{\mathrm{bg}} <$ 100 V.

The fact that the signal is quadratic in the current may be exploited to remove several potential sources of measurement error.  The $I^2$ dependence means that an AC current at frequency $\omega$ produces a transverse potential at frequency $2 \omega$.  We use a lock-in amplifier to measure the potential at $2 \omega$ while filtering out potentials at $\omega$.  A Hall potential due to a DC magnetic field, such as that of the earth, will appear at $\omega$ and thus can be safely ignored.  The potential drop due to the longitudinal electric field component within the wire, $E_\theta$, likewise occurs at $\omega$, so we need not worry about imperfect alignment of transverse measurement leads.  By checking that $\Delta V_{\mathrm{geom}}$ is proportional to $I^2$, we ensure that any $2 \omega$ harmonics in the current do not contribute to the signal.  Fluctuations in the conductor's resistivity due to Joule heating give rise to \textit{longitudinal} oscillations in the potential that occur at $3 \omega$ and higher harmonics, but not at $2 \omega$. (See SI for details.)

Two extraneous sources of a $2 \omega$ signal are due to (i) the Hall voltage from a current-induced magnetic field and (ii) the Seebeck effect. In the SI we describe how we have minimized their contribution so that they do not affect our results.

We have measured the potential difference, $\Delta V_{\mathrm{geom}}$, across the curved wires in our samples. In all the samples, without any current-annealing and at $V_{\mathrm{bg}} = 0$, $\Delta V_{\mathrm{geom}}$ is positive, corresponding to positive charge carriers. This is what is expected in graphene on a SiO$_{2}$ substrate~\cite{Novoselov2004,Martin2008,Wang2010}.  
We fit the measured transverse potentials to a power $\beta$ of the current amplitude, $| \Delta V_{\mathrm{geom}} | \propto I^\beta$, and we find $\left< \beta \right> = 2.00 \pm 0.11$.  This confirms that the measured potential rises quadratically with the current, as shown in Fig.~2A. There is significant scatter in the signal magnitude between samples as shown in Fig.~2B. For a driving voltage $\mathcal{E} =$ 1.00 V, the current is $I \approx 370 \pm 130$ $\micro$A and the average magnitude of the signal is $\left<\Delta V_{\mathrm{geom}}\right> \approx$ 0.46 mV.  Using Eq.~\ref{eq:V_trans} and averaging over the samples, we measure $\left< m / (n^2t^2 q^3) \right> \approx 5 \times 10^{-6}$ kg m$^4$ C$^{-3}$.

\begin{figure}
    \centering
    \includegraphics{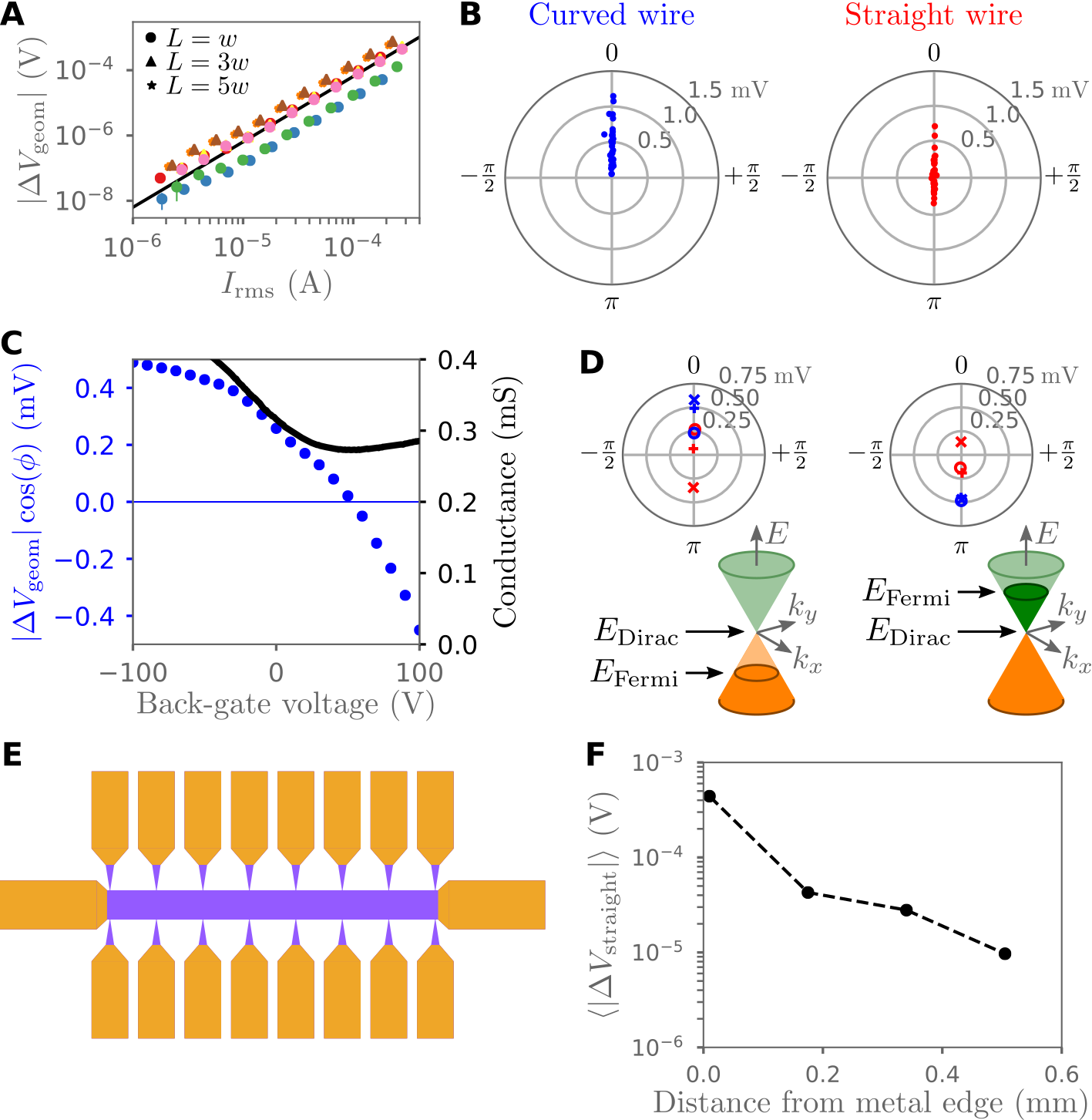}
    \caption{
    Measured transverse potentials in graphene wires.
    (\textbf{A}) Signal across curved wire versus current.  Colors correspond to different samples, each with $r_{\mathrm{in}} =$ 10 $\micro$m, $r_{\mathrm{out}} =$ 100 $\micro$m, and graphene measurement leads of length $L \geq$ 90 $\micro$m.  Black line, average of power-law fits to individual data sets, has slope = 2.0.
    (\textbf{B}) Signals from curved (blue) and straight (red) wires in 34 samples at an average current $\left< I_{\mathrm{rms}} \right> \approx$ 370 $\micro$A.  The radius and angle represent the magnitude and phase of the measurement.  Error bars are smaller than data markers.  
    (\textbf{C}) Signal across a curved wire (blue) and circuit's conductance (black) versus $V_{\mathrm{bg}}$ after current-annealing.
    (\textbf{D}) Measurements from curved wires when the Fermi level is below (left) or above (right) the Dirac point, controlled by changing $V_{\mathrm{bg}}$ after current-annealing.  Symbols correspond to different samples.  For a curved wire phase $\phi = 0$ indicates positive charge carriers. 
    (\textbf{E}) Long wire design with 8 pairs of measurement leads.
    (\textbf{F}) Signal versus distance from the metal edges at either end of the straight wire, averaged across back-gate-voltage sweeps from -100 V to +100 V in four samples.
    }
    \label{fig:fig2}
\end{figure}

After current annealing, we can apply a back-gate voltage $V_{\mathrm{bg}}$ to the sample and move the Fermi level to the other side of the Dirac point where the charge of the carriers has the opposite sign.  We find that $\Delta V_{\mathrm{geom}}$ changes sign at a back-gate voltage close to that of the conductance minimum, as shown in Fig.~2C.  This confirms that this measurement determines the sign of the charge carriers using only geometry. (This behavior shows hysteresis with the direction of the $V_{\mathrm{bg}}$ sweep, as is typical of electrical properties of graphene~\cite{Wang2010}.)  Rather than showing a singularity, $\Delta V_{\mathrm{geom}}$ passes smoothly through zero near the Dirac point.  This is consistent with prior observations that in graphene samples near the Dirac point, inhomogeneities and defects make the graphene behave as a random assortment of electron and hole puddles rather than as a uniform material with $n \approx 0$~\cite{Martin2008,Lee2008,Blake2009,Zhang2009}.

Because our samples are thin, the current $I$ is small even though the current density $j$ is very large. Therefore, the magnitude of the Hall-effect contribution from a current-induced magnetic field should be (see Fig. S1) at least 20 times smaller than the signal we observe and the prediction of Eq.~\ref{eq:V_trans}.   We have also checked (see SI) that $\Delta V_{\mathrm{straight}}$ is not due to the Seebeck effect by varying the length of the graphene voltage-measuring leads.

The sign of the potential difference, $\Delta V_{\mathrm{straight}}$, between the two sides of the straight wire, is not picked out by the curvature of the wire.  However, we still find a signal whose phase is either $\phi = 0$ or $\phi = \pi$, as shown in Fig.~2B.  The average signal over many samples $\left<\Delta V_{\mathrm{straight}}\right> \approx 0$, but in a single sample, the magnitude of the signal in the straight section can be comparable to, but usually smaller than, that in the curved section.  We find $\left< |\Delta V_{\mathrm{straight}}| \right> \approx 0.35 \left< |\Delta V_{\mathrm{geom}}| \right>$. As we sweep $V_{\mathrm{bg}}$ in a sample that has been current-annealed, $\Delta V_{\mathrm{straight}}$ changes sign (just as does $\Delta V_{\mathrm{geom}}$), as shown in Fig.~2D.

We have measured $\Delta V_{\mathrm{straight}}$ from a long straight wire with 8 pairs of electrodes across it, illustrated in Fig.~2E.  We average over four long-wire samples and find that the magnitude of $\Delta V_{\mathrm{straight}}$ falls off with $d$, the distance from where the graphene makes contact with the metal source and drain as shown in Fig.~2F.  The potential drops by nearly two orders of magnitude from near the leads, where $d \approx$ 10 $\micro$m, to the center, where $d \approx$ 500 $\micro$m.

It is unexpected that there is a significant signal across a straight section of wire.  The sign of this voltage varies from place to place along the wire and in different samples and its magnitude is smaller than, but comparable to, that in the curved section.  We conclude that the reason for this behavior is that the current paths are not homogeneous.  The signal observed in the straight sections is thus due to the current taking a meandering path along the wire.

To rationalize the decrease of $\Delta V_{straight}$ from the ends of the wire, we assume that the current is injected from the Ti/Au leads into the graphene at localized points; the current then fans out as it moves down the wire. This is consistent with studies showing that metal contacts introduce inhomogeneous doping~\cite{Blake2009}, and that contact may be poor due to surface impurities on the graphene~\cite{Martin2008,Allain2015}.  Photocurrent mapping of graphene transistors has revealed irregular electrostatic potential landscapes, including at the metal contacts and along the edges~\cite{Lee2008}.  As the current flows down the wire, it expands to fill more of the wire's width.  Because the signal is quadratic in the current density, for fixed total current, the smaller the width, the larger will be the signal.  

This interpretation also offers an explanation for the large measured magnitude of $\left<\Delta V_{\mathrm{geom}}\right>$.  If the mass in Eq.~\ref{eq:V_trans} is taken to be the bare mass of the electron with charge $q=e$, our data suggest a carrier density $n \approx 10^{12}$ cm$^{-2}$, which is a tenth the value expected for our doping level~\cite{Novoselov2004}.  However, if the current paths are not given by the wire width, $w$, but rather by the heterogeneity of the current path, then we can account for the observed large value of $\left<\Delta V_{\mathrm{geom}}\right>$ by using a smaller width in Eq.~\ref{eq:V_trans}.

There is considerable evidence that currents in thin metal films~\cite{Zhang2006Au,Yoneoka2012} and in two-dimensional conductors such as graphene are not uniform throughout the wires.  In graphene this has been ascribed to scattering at grain boundaries~\cite{Tsen2012}, as well as to charge puddles~\cite{Martin2008,Zhang2009} and local strains~\cite{Pereira2009,Guinea2010}.  The measurement of a transverse $2\omega$ signal, $\Delta V_{\mathrm{straight}}$, is an elegant probe of the tortuous current path.

We have demonstrated a transverse voltage across a current-carrying wire due to geometry alone.  In the classical Hall effect, a magnetic field curves the paths of charge carriers inside a straight wire so that charges accumulate on the wire edges transverse to the current. In the geometric analog, we do not bend the paths of the carriers but instead bend the conductor itself to create a purely geometric effect.  The observed signal is consistent with a prediction from elementary mechanics and electrodynamics.  

We observe signals even in straight wires.  Although the wires themselves are straight, the internal current paths are not.  Just as for curved wires, charge distributions are necessary to confine currents to any heterogeneous path.  The nonlinear transverse voltage offers a novel technique for studying such heterogeneities.

Recent work~\cite{ma2019} has found a nonlinear Hall effect due to an induced Berry curvature~\cite{Xiao2010} in bilayer conductors. Such an effect is not expected to occur in a single-layer material such as graphene and is not predicted to depend on the wire curvature.  Our purely geometric effect can contribute to the signals found in those experiments and in turn those effects, if present, could masquerade as a geometric effect.  

Low-temperature quantum Hall effects arise due to time-reversal-symmetry breaking in a magnetic field. In the presence of quantum interactions, the magnetic Hall effect becomes particularly remarkable; it would be interesting to consider if any striking quantum effects can be observed due to geometry alone.

\bibliographystyle{unsrt}
% \bibliography{main}

\section*{Acknowledgments}

We are particularly grateful to Kh\'{a}-\^{I} T\^{o} who worked on the early stages of this project and to Lujie Huang who gave important advice about nanofabrication with graphene.  We thank Jiwoong Park's group (J. Park, K.-H. Lee, J.-U. Lee, and P. Poddar) as well as G. Koolstra, N. Earnest, S. Chakram, and F. Tang for technical assistance.  We thank  P.B. Littlewood, C. Panagopoulos, and D.T. Son for helpful discussions.
% \textbf{Funding:}
 This work was supported by the NSF MRSEC Program DMR-1420709 and NSF DMR-1404841 and used the Pritzker Nanofabrication Facility, supported by NSF ECCS-1542205.
 % \textbf{Authors contributions:}
 % N.B.S. prepared samples, conducted measurements, and analyzed the data. 
 % S.R.N. and D.I.S. planned and conceived the study.
 % N.B.S. and S.R.N. wrote the manuscript.
 % \textbf{Competing interests:} The authors declare no conflicts of interest.
 % \textbf{Data and materials availability:} All data is available upon request.

\clearpage

\end{document}

% Double-space the manuscript.
\baselineskip24pt

\maketitle
\pagestyle{plain} % remove running title

\section{Materials and Methods}

\subsection{Circuit fabrication}

We begin with CVD graphene that has already been transferred by the manufacturer (Graphenea) to a doped silicon wafer with a \unit{300}{\nano \meter} SiO$_2$ gap, and diced to \unit{10}{\milli \meter} $\times$ \unit{10}{\milli \meter} chips in a class 1000 cleanroom.  To improve adhesion between the graphene and the oxide substrate, we anneal the samples at \unit{300}{\degreecelsius} in nitrogen for at least 6 hours using a Gemstar ALD (Arradiance).

We use the cleanroom in the Pritzker Nanofabrication Facility at the University of Chicago for our photolithography procedure.  Our first round of photolithography is for the metal portions in our circuit designs.  We first spin LOR 3A (MicroChem) onto the graphene at 500 rpm for 10 s and then 3000 rpm for 45 s, for an undercut during development.  We then bake the sample at \unit{180}{\degreecelsius} for \unit{5}{\min}.  We next spin AZ 1512 photoresist (Clariant) onto the sample at 500 rpm for 10 s and then 4500 rpm for 45 s, and bake at \unit{115}{\degreecelsius} for \unit{1}{\min}.  We use a Heidelberg MLA150 Direct Write Lithographer to expose the pattern for our metal pads onto the chip, using a \unit{405}{\nano \meter} laser and a dose of 100 mJ/cm$^2$.  We develop the photoresist in AZ 300 MIF (Clariant) under gentle agitation for \unit{1}{\min} and then transfer to deionized water for \unit{1}{\min}.  We immediately dry the sample with nitrogen.

We use electron-beam evaporation (Angstrom Nexdep) to deposit a layer of metal onto the substrate for the electrode pads.  We first deposit \unit{2}{\nano \meter} of titanium (0.5 \AA/s) and then \unit{50}{\nano \meter} of gold (1.0 \AA/s).  We perform lift-off by submerging overnight in AZ NMP (Clariant) at room temperature.  The next morning, we rinse the sample with acetone (Fisher/VWR) and isopropyl alcohol (Fisher/VWR) and then dry it with nitrogen.  We are careful to prevent the chip from drying out while it is exposed to acetone.

We next perform another round of photolithography to pattern the graphene itself.  We spin poly(methyl methacrylate) (495 PMMA A 4, MicroChem) onto the substrate at 500 rpm for 10 s and then 4000 rpm for 60 s.  We bake the sample at \unit{145}{\degreecelsius} for 5 min.  Next we spin AZ 1512 photoresist onto the sample again at 500 rpm for 10 s and then 4500 rpm for 45 s.  We bake the sample at \unit{115}{\degreecelsius} for \unit{1}{\min}.  Using the direct write lithographer once again, we align to our previous pattern and then expose an inverted pattern for the graphene and metal portions of the circuits, using the \unit{405}{\nano \meter} laser and a dose of 100 mJ/cm$^2$.  During this step, we expose the areas where the graphene will ultimately be \textit{removed}.  We develop the photoresist with the same steps that we use in our first round of photolithography.

We use oxygen plasma etching (YES CV200 Oxygen Plasma Strip / Descum System) to remove the exposed PMMA layer and to remove the graphene once it is exposed.  Using 50 sccm of oxygen, we etch at \unit{400}{\watt} for 80 s.  We inspect under an optical microscope afterward to determine whether the graphene has been completely removed from the exposed areas.  If it has not, we etch for 20 s or 40 s longer.  Finally, we strip off the unexposed photoresist and PMMA by submerging the sample vertically in acetone for at least 6 hours.  Afterward, we rinse the sample with isopropyl alcohol and dry with nitrogen.

We inspect each device on the chip under an optical microscope to make sure that the metal and graphene regions are intact and well-aligned.  We vacuum-seal the chips for storage under low vacuum once fabrication is complete.

\subsection{Current annealing}

As initially fabricated, the samples are highly doped; the Fermi level is far from the Dirac point. Under these conditions it is impractical to move the Fermi level to the other side of the Dirac point by applying a back-gate voltage. However, we can current-anneal [17] the samples by injecting a current of $I \approx$ \unit{3}{\milli \ampere} ($\left<j\right> \approx 10^7$ A/cm$^2$). 
To do this, we apply \unit{10.0}{\volt} between the source and drain of the device for at least 2.5 hours and typically overnight, while the sample is exposed to air.
Once this is done, the back-gate voltage corresponding to the graphene conductance minimum typically falls within the range of 0 to \unit{+100}{\volt}, so we can access the other side of the Dirac point by using a back-gate voltage less than \unit{+100}{\volt}.

\subsection{Electrical measurements}

We perform electrical measurements on the graphene circuits using a probe station where the samples are exposed to air at room temperature.  We use a SR830 lock-in amplifier (Stanford Research Systems) for transverse potential measurements at frequencies between \unit{10}{\hertz} and \unit{250}{\hertz}.  To measure the resistance of the graphene wires, we use a Model SR570 Low-Noise Current Preamplifier (Stanford Research Systems) and a BNC-2110 Shielded Connector Accessory (National Instruments).  We control the back-gate voltage using a HP 6827A Bipolar Power Supply/Amplifier.

\section{Sources of error in measurement}

\subsection{Joule heating oscillations in longitudinal potential}

By using a lock-in amplifier to measure the potential at $2 \omega$, we filter out Ohmic potential drops along the wire as a possible source of error because they occur at $1 \omega$.  However, one might ask whether Joule heating could lead to oscillations in the longitudinal potential that could appear at $2 \omega$.  The resistance $R$ of a wire will fluctuate with temperature $T$ relative to its steady-state resistance $R_0$ and temperature $T_0$ as

\begin{equation}
    R(T) \approx R_0 \left[ 1 + \alpha \left( T - T_0 \right) \right],
    \label{eq:R_vs_T}
\end{equation}

\noindent where $\alpha$ is the material's temperature coefficient of resistance.  The temperature fluctuates in time $t$ with power dissipation $P$ due to Joule heating,

\begin{equation}
    T(t) \approx T_0 + \beta P(t),
    \label{eq:T vs P}
\end{equation}

\noindent for some coefficient $\beta$.  The power dissipation, in turn, is related to the resistance by

\begin{equation}
    P(t) = I^2 R(T).
    \label{eq:P_vs_R}
\end{equation}

Substitution of these relationships into Ohm's Law, $\mathcal{E} = IR$, leads to higher-order terms in the longitudinal potential drop:

\begin{equation}
    \mathcal{E} \approx IR_0 + \alpha \beta I^3 R_0^2 + ...
    \label{eq:Ohm_corrections}
\end{equation}

\noindent With AC current at frequency $\omega$, the $I^3$ term in Eq. \ref{eq:Ohm_corrections} affects only odd harmonics, so there is no interference with a measurement at $2 \omega$.

\subsection{Seebeck effect}

Joule heating can raise the temperature of one side of the wire more than the other if, for example, the current density, $j(r)$, depends on radius or is otherwise not homogeneous.  This can induce a Seebeck voltage \textit{across} the wire where the graphene makes contact with a conducting lead of a different material.  The temperature at any point in the graphene wire will oscillate due to Joule heating:

\begin{equation}
\begin{aligned}
    T(\mathbf{r},t) &\approx T_0(\mathbf{r}) + \alpha(\mathbf{r}) j^2(\mathbf{r}, t) \\
    & \approx T_0(\mathbf{r}) + \alpha(\mathbf{r}) j_0^2(\mathbf{r}) 
      \left[ \frac{1 - \cos( 2 \omega t ) }{2}  \right],
    \label{eq:temp_joule}
\end{aligned}
\end{equation}

\noindent where the function $\alpha(\mathbf{r})$ depends on the frequency as well as the material's specific heat capacity and resistivity, which could vary spatially.  The dependence on $j^2$ means that the local temperature oscillates at $2 \omega$ in our experiments.  The temperature \textit{difference} $\Delta T$ between two points in the wire, $\mathbf{r}_1$ and $\mathbf{r}_2$, will oscillate at $2 \omega$ with an amplitude that depends on the differences between the current density amplitudes $j_0(\mathbf{r})$ and the local values of $\alpha (\mathbf{r})$ at those two points in the wire:

\begin{equation}
    \Delta T(\mathbf{r}_1, \mathbf{r}_2, t) \approx T_0(\mathbf{r}_1) - T_0(\mathbf{r}_2) + 
    \left[\alpha(\mathbf{r}_1) j_0^2(\mathbf{r}_1) 
    - \alpha(\mathbf{r}_2) j_0^2(\mathbf{r}_2) \right]
      \left[ \frac{1 - \cos( 2 \omega t ) }{2}  \right].
    \label{eq:temp_diff}
\end{equation}

This will create a transverse signal at $2 \omega$. For this reason we keep our metal measurement leads a distance of at least one wire width, $w$, away from the conducting path of the wire, so that $j(\mathbf{r}_1) \approx j(\mathbf{r}_2) \approx 0$.  The rest of the measurement lead is made of the same graphene as the wire itself. By moving the point of connection away from where there is any temperature variation, we can minimize the influence of this Seebeck effect.  We have varied the length $L$ of the graphene leads between $1w$ to $5w$ and have not found any systematic variation of the $2\omega$ signal with their length.

\subsection{Hall voltage due to a current-induced magnetic field}  

The most important source of an extraneous signal at $2 \omega$ is the Hall effect due to the magnetic field generated by the current itself.  An oscillating electric current at $\omega$ through the circuit contributes to a time-varying magnetic field $\mathbf{B}(\mathbf{r}) \propto I$. 
Eq.~1 (main text) shows that this produces a Hall potential that is quadratic in the current and which therefore appears at $2 \omega$.  We note that the magnetic field created by such a current is proportional to the \textit{current}, while $\Delta V_{\mathrm{geom}}$ is dependent only on the \textit{current density} (or $I/t$) as shown in Eq.~3 (main text). Thus, we can minimize this extraneous source of a $2\omega$ signal by reducing the thickness, $t$, of the conducting wire; by making the wire thin, we reduce the current without changing the current density.  By using atomically thin, monolayer graphene, we have minimized the thickness, $t$.

We can express the magnetic field as

\begin{equation}
    \mathbf{B}(\mathbf{r}) = \frac{\mu_0 I C_{\mathrm{self}}(\mathbf{r})}{r}.
    \label{eq:V_self_Hall}
\end{equation}

\noindent The dimensionless function $C_{\mathrm{self}}(\mathbf{r})$ depends on how the entire circuit is arranged.  We include the $r$ in the denominator of Eq.~\ref{eq:V_self_Hall} because the denominator must have units of length and is thus set by a length scale in the system.

\subsubsection{Estimation of effect size for graphene circuits}

The Hall potential between our measurement leads is determined by the component of the magnetic field perpendicular to the plane of the circuit pattern.  The largest contribution to this component of the magnetic field comes from the wires that are actually in that plane --- the wires on the substrate itself.  Therefore, we choose $r \equiv w = ( r_{\mathrm{out}} - r_{\mathrm{in}})$ for the denominator, since $w$ is the length scale that characterizes our graphene wire within the plane, and $w$ is also the approximate distance that the curved wire in our design is offset from the straight portions of the circuit pattern.

We next need to make an assumption about the value of the prefactor $C_{\mathrm{self}}(\mathbf{r})$ at the locations in our circuit where we measure transverse potentials.  We note that in the case of a point that is a distance $w$ away from an infinite wire carrying a current $I$, Amp\`{e}re's Law tells us that $C_{\mathrm{self}} = 1 / (2 \pi) \approx 0.16$.  In order to ensure that we are not underestimating the magnitude of $\mathbf{B}$, we set $C_{\mathrm{self}} = 1$.  We can then substitute for $\mathbf{B}$ to estimate the self-induced Hall potential:

\begin{equation}
    V_{\mathrm{Hall}} = C_{\mathrm{self}} \frac{\mu_0 I^2}{n t q (r_{\mathrm{out}} - r_{\mathrm{in}})} \le \frac{\mu_0 I^2}{n_{\mathrm{2D}} q (r_{\mathrm{out}} - r_{\mathrm{in}})},
    \label{eq:V_self_Hall_bound}
\end{equation}

\noindent where $t$ in this equation represents the thickness of the wire out of plane.  The two-dimensional carrier density $n_{\mathrm{2D}} = n t$ is given by

\begin{equation}
    n_{\mathrm{2D}} = \frac{\varepsilon_0 \varepsilon V_{\mathrm{bg}}}{d q},
    \label{eq:carrier_density}
\end{equation}

\noindent as described by Novoselov \textit{et al} [16].  Here $\varepsilon_0$ and $\varepsilon$ are the permittivities of free space and SiO$_2$, $V_{\mathrm{bg}}$ is the back-gate voltage relative to the Dirac point, and $d$ is the thickness of the SiO$_2$ layer.

For any given value of the current, we find that the theoretical value of $\Delta V_{\mathrm{geom}}$ (assuming uniform material properties in the graphene) is more than an order of magnitude larger than $\Delta V_{\mathrm{Hall}}$, as shown in Fig.~S1.  For this calculation we assume the graphene is sufficiently doped that the charge neutrality point is 100 V away in back-gate voltage.

\begin{figure}
    \centering
    \includegraphics{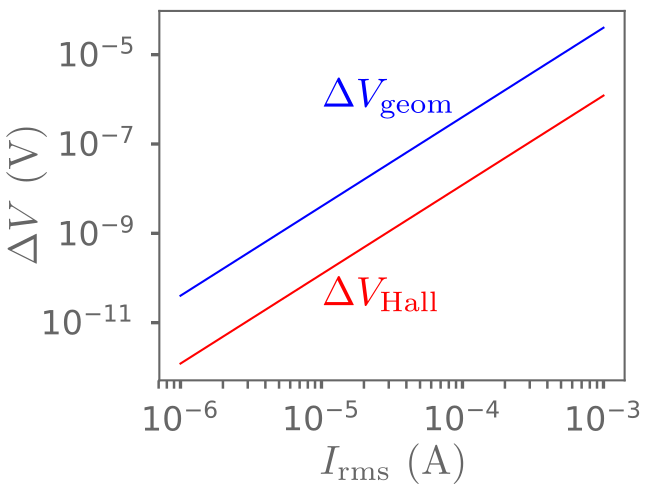}
    \caption{Predicted values of transverse potentials due to centripetal acceleration (blue) and due to the Hall effect from the magnetic field generated by the current itself (red).
    % Plot is from plot_geom_v_hall_20190117-1.py
    }
    \label{fig:figS1}
\end{figure}

\subsubsection{Alternative options}

Minimizing the wire thickness $t$ is not the only means of minimizing the current-induced Hall effect relative to $\Delta V_{\mathrm{geom}}$.  One option is to use a superconductor, which will expel any magnetic fields as long as they do not exceed the critical field of the superconductor.  This effectively places an upper bound on the current or current density that one may use for the measurement in the superconductor, and one must ensure that the noise floor of the lock-in amplifier does not exceed the expected size of the signal $\Delta V_{\mathrm{geom}}$ when using that amount of current.

Another option is to intentionally design the circuit's geometry to minimize the function $C_{\mathrm{self}}(\mathbf{r})$ at the location of the curved wire, as a way to minimize $\Delta V_{\mathrm{Hall}}$ in Eq.~\ref{eq:V_self_Hall_bound}.  One way to accomplish this would be a two-layer circuit design.  In the first layer, a thin, flat wire follows a path through a curved section and a straight section, similar to the graphene circuit design that we have used.  At one end of the pattern, however, the wire could instead connect to a second conducting layer, directly on top of the first, with a thin insulating spacer, such that the wire then traces out an identical path but such that the current will travel through the second layer in the opposite direction.  The contribution of both layers to the out-of-plane magnetic field would be minimized, for the same reason that the magnetic field is minimized outside of a coaxial cable or a pair of twisted cables that carries current in both directions.